\documentclass[a4paper,12pt]{article}
\pdfoutput=1

\usepackage{test} 
\usepackage{booktabs} 
\usepackage{caption}
\usepackage{appendix}
\usepackage{hyperref}
\usepackage{longtable}
\usepackage{lscape}
\hypersetup{colorlinks,citecolor=blue,urlcolor=magenta}
\usepackage{doi}

\usepackage[style=authoryear-comp,
sorting=nyt,
dashed=false, 
maxcitenames=2, 
maxbibnames=99, 
uniquelist=false,
uniquename=false,
giveninits=true, 
natbib, 
date=year 
]{biblatex}

\AtBeginRefsection{\GenRefcontextData{sorting=ynt}}
\AtEveryCite{\localrefcontext[sorting=ynt]}

\DeclareFieldFormat{pages}{#1} 
\renewbibmacro{in:}{\ifentrytype{article}{}{\printtext{\bibstring{in}\intitlepunct}}} 

\usepackage[inline,shortlabels]{enumitem}
\setlist[enumerate,1]{label=(\roman*)}

\usepackage[margin = 1.25in]{geometry}
\usepackage{setspace}
\title{Pareto's Limits: Improving Inequality Estimates in America, 1917 to 1965}
\author{Vincent Geloso\thanks{Department of Economics, George Mason University. Email: \href{mailto:vgeloso@gmu.edu}{vgeloso@gmu.edu}.} \and Alexis Akira Toda\thanks{Department of Economics, Emory University. Email: \href{mailto:alexis.akira.toda@emory.edu}{alexis.akira.toda@emory.edu}.}}

\numberwithin{equation}{section}

\usepackage{bbm}
\renewcommand{\hat}{\widehat} 

\begin{document}
\maketitle

\begin{abstract}

American income inequality, generally estimated with tax data, in the 20th century is widely recognized to have followed a U-curve, though debates persist over the extent of this curve, specifically regarding how high the peaks are and how deep the trough is. These debates focus on assumptions about defining income and handling deductions. However, the choice of interpolation methods for using tax authorities' tabular data to estimate the income of the richest centiles---especially when no micro-files are available---has not been discussed. This is crucial because tabular data were consistently used from 1917 to 1965. In this paper, we show that there is an alternative to the standard method of Pareto Interpolation (PI). We demonstrate that this alternative---Maximum Entropy (ME)---provides more accurate results and leads to significant revisions in the shape of the U-curve of income inequality.

\medskip
		
		
\medskip
		

\end{abstract}

	
\section{Introduction}

In the early 1950s, Simon Kuznets unveiled a comprehensive set of estimates detailing income inequality in America from 1917 to 1948 \citep{Kuznets1953}. These estimates were derived from multiple editions of the \textit{Statistics of Income} produced by the Internal Revenue Service (IRS). Kuznets' findings highlighted a significant leveling of inequality, particularly after 1929. Subsequent seminal articles and books have not only confirmed this observation but have also demonstrated that income inequality in America followed a U-curve pattern since the early 20th century \citep{paglin1975measurement, williamson1980american,goldin1992great,smiley2000note, PikettySaez2003,PikettySaez2007,lindert2017unequal}. This pattern demonstrates high inequality in the 1910s and 1920s, a leveling off around 1940, persistence until the late 1970s, and a sharp increase in inequality thereafter. This U-curve serves as backdrop for multiple debates concerning the ebbs and flows of the welfare state, the taxation level on top incomes, the role of labor unions in reducing inequality, political institutions, and social trust, among others. 

Few dispute the now-stylized fact of the U-curve. However, there are debates regarding the timing of the changes, their magnitude, and the overall level (independent of timing). The debate is divided in two periods corresponding to the curve's left and right sides. On the left side---the period from 1917 to the 1960s---some argue that inequality began high, plateaued, and then collapsed during WWII due to the expansion of the tax base and higher tax rates on top incomes. A low plateau was reached by the early 1950s. That low plateau is a ``golden age'' with ``three decades of rapid growth'' that benefited mostly the bottom rungs on the income ladder \citep[p. 207]{lindert2017unequal}. The result is described as the ``great leveling" \citep*{lindert2017unequal}. On the right side of the curve, we start at the low plateau of the 1950s that continued until the late 1970s, followed by a continuous increase back to the level of the 1920s today \citep*{PikettySaez2003, PikettySaez2007, piketty2018distributional,lindert2017unequal}.

These arguments have been countered on both sides of the curve. On the curve's left side, others argue that better datasets indicate a lower level of inequality with a more gradual leveling off and no precipitated decline during the 1940s as a result of high marginal tax rates on top incomes.  \citep*{smiley2000note, geloso2020great, GelosoMagnessMooreSchlosser2022}. They also argue that the plateau was not as deep as claimed because the ratio of corporate to personal income tax rates caused people to shift income between legal entities \citep*{reynolds2006income, mechling2017piketty, AutenSplinter2024}.  Improvements to the dataset on the right side of the curve suggest a large increase during the 1980s, which then stabilized at a higher plateau that remains below the 1920s level \citep*{armour2014levels, AutenSplinter2024, larrimore2021recent}. The overall level is also lower. For them, the pattern is more like a tea-saucer than a pronounced U-curve. 

In this paper, we argue that the debates focus primarily on data quality while overlooking a crucial methodological choice that affects the curve's left side. Before 1966, the tax data used to estimate income inequality did not come from micro-files; instead, researchers rely on tabular data reported by the IRS. These tables provide the number of filers by income classes (e.g., \$0 to \$1,000, \$1,000 to \$2,000, etc.) and the total income reported for each class. To estimate the average income of the top 10\%, 5\%, 1\% (etc.) of the population, an interpolation method is necessary---something that is not required with micro-files. The method universally employed is the Pareto Interpolation (PI) method, a parametric density estimation technique that approximates the average income of the richest percentiles of the population. 

We leverage a new approach on nonparametric density estimation---Maximum Entropy (ME)---by income class developed by \citet*{LeeSasakiTodaWang2024JOE} to assess whether the undisputed use of Pareto interpolation is warranted. When micro-data are available to compare (\ie, after 1965), we find that ME is closer to true values than PI. This is particularly true when the IRS tables report few income classes (meaning that PI is less precise as the number of income classes reported falls). Using ME instead of PI alters the evolution of income inequality from 1917 to 1965 (\ie, the period when one is forced to rely on tabular summary data). The level of inequality is not noticeably different between methods for the 1920s and 1930s. However, because of changes in the way the IRS reported and collected the data, larger and directionally consistent errors (with PI underestimating inequality relative to ME) start appearing in the 1940s. This continues until the 1960s. This has a serious implication in that the ``golden age'' of equality (\ie, the end point of the ``great leveling'') had higher inequality than commonly understood. 

This attenuates the U-curve pattern. It pushes in the direction of the claims of scholars who have been pushing against the numbers advanced by \citet{PikettySaez2003, PikettySaez2007,piketty2018distributional} for a very pronounced U-curve. For example, \citet{GelosoMagnessMooreSchlosser2022} argue that the U-curve looks more like a ``shallow tea saucer'' when the data is corrected for the issues they raised. Our findings suggest a shallower saucer. 

Our paper is divided as follows. Section \ref{sec:method} briefly explains the income tax data used to estimate inequality pre-1965 and then carries on to explain the two different interpolation methods. Section \ref{sec:accuracy} compares the effectiveness of the two methods against micro-files data when available. Section \ref{sec:result} presents the results of the ME estimation and discusses the importance of our results.

\section{Methodology and data}\label{sec:method}

The computation of top income shares pre-1965---when we are forced to rely on tabular data rather than micro-data that span the entire population---is quite laborious.\footnote{As \citet{PikettySaez2007} point out, ``micro-files were constructed annually since 1966'' (p.~167). ``For the 1966–99 period, the series were computed directly from the IRS micro-files'' whereas for the 1913–65 period ``the series were estimated from the published IRS tables using the Pareto interpolation technique'' (p.~195).} To understand, we can break the computation into two sets of independent steps. The first speaks to adjusting the data \textit{before} making the choice of interpolation method. To the best of our knowledge, \textit{all} the debates regarding pre-1965 estimates center on this set. The second set consists in the choice of interpolation method (the object of this paper) to pinpoint where the different fractiles (\eg, top 10\%, 5\%, 1\% etc.) are located. We first briefly explain how the data is prepared for the period 1917 to 1965 with the aim to show that whatever choices being made in that respect will not affect the choice of interpolation method. 

\subsection{Preparing the data}\label{subsec:method_data}

Before proceeding with our discussion of interpolation methods for the period from 1917 to 1965, we must do a digression to explain more about the IRS data taken from the \textit{Statistics of Income} and how they are used. This is because there are some difficulties of importance with the data. First, the units in the IRS tables are for tax returns, not households. This means that we are not expressing income per household but rather income per tax unit.\footnote{There are more than one tax units per household on average.} 

Second, pre-1940, it is rare to find more than 10\% of tax units who filed tax returns. There are many years with too few filers to estimate the top 10\% and top 5\% income shares. \citet{PikettySaez2007} dealt by applying the ratio of married to single filers from income brackets above the minimum tax eligibility threshold (generally those below \$5,000) in all pre-1940 years (pp.~195--6).\footnote{This is because married filers had higher exemptions than single households.} This ratio allowed them to increase the size of the lower income classes where too few returns were filed by married individuals. The adjustment was made using the 1942 ratio of married to single filers, derived from a sample of 1040A for that year, as a proxy for tax filing conditions in the previous two decades. \citet{GelosoMagnessMooreSchlosser2022} pointed out that this correction was flawed because 1942 was a war year, during which many men were enlisted, potentially skewing the adjustment. They also noted that a similar sample existed for 1941 (\ie, a peacetime year) that was larger than the one used by \citeauthor{PikettySaez2007} (516,000 returns versus 455,000 returns) and covered all IRS collection districts. Using that sample, they found that the multipliers were generally half those calculated by \citeauthor{PikettySaez2007}.

Third, prior to 1944, the IRS reported income classes by net income (\ie, gross income minus deductions). As a result, all the reported totals must be converted into adjusted gross income (AGI). For \citet{PikettySaez2007}, the conversion was made using fixed ratios according to each fractile $p$ (\eg, top 10\%, 5\%, 1\%, etc.). In contrast, \citet{GelosoMagnessMooreSchlosser2022} collected deductions data from a wide variety of governmental sources to make the adjustments. They found that the ratios used by \citet{PikettySaez2007} were unaligned with the actual data.\footnote{In their \texttt{comp1398.xls} file, \citet{PikettySaez2007} state, ``je me suis content\'e de reprendre des valeurs raisonnables \`a partir des \texttt{tabcomp} reproduits sur la feuille \texttt{Comp1665},'' meaning, ``I was content in taking reasonable values inferred from \texttt{tabcomp} [income breakdowns] on sheet \texttt{Comp1665}.'' No justification across all their works is given for the ``reasonableness'' of these inferences.}

Fourth, there is the issue of the income denominator. \citet{PikettySaez2003,PikettySaez2007} estimated fiscal income pre-1944 by using 80\% of total personal income minus transfers. After that, they assigned 20\% of the average income of tax filers to non-filers to generate the income denominator. \citet{GelosoMagnessMooreSchlosser2022} used a line-by-line reconciliation of national accounts with tax data. 

Overall, there are substantial differences in the resulting estimates, with \citet{GelosoMagnessMooreSchlosser2022} showing consistently lower levels of income inequality than \citet{PikettySaez2003,PikettySaez2007}, with stronger leveling during the Great Depression and weaker leveling during World War II.

We use both datasets of \citet{PikettySaez2003} and \citet{GelosoMagnessMooreSchlosser2022}, for three reasons.\footnote{We should note that we are benchmarking ourselves against the \citeyear{PikettySaez2003} work of \citeauthor{PikettySaez2003} rather than their updated figures on their websites. This is because they shared only the codes, procedure and data for these initial estimates. Since then, they have made revisions to their methods and they have never shared how the minor differences between the \citeyear{PikettySaez2003} work and later work appeared.} First, both rely on the same method (\ie, Pareto interpolation) and neither considered other interpolation methods pre-1965. Second, none of the debated steps to adjust the data affect the interpolation method: the choice of adjustments to prepare the data is independent from the choice of interpolation methods. Third, it avoids the debate over which dataset is correct. Both should be affected in the same way by the choice of the interpolation method, which means our modifications are neutral with respect to existing preferences regarding the choice of estimates.

\subsection{Pareto interpolation (PI) method}\label{subsec:method_PI}

To help visualize the interpolation method, Table \ref{tab:PSData} below contains an abbreviated version of the \textit{Statistics of Income} for 1920. Let the income classes be indexed by $k=1,\dots,K$, in the order from highest to lowest (Column (1), where $K=37$). Let the income thresholds for these classes be denoted by $\infty=t_0>t_1>\dots>t_K$ (Column (2)), so income class $k$ has income in the range $I_k\coloneqq [t_k,t_{k-1})$. Let $n_k$ be the number of observations with income at least $t_k$ and $S_k$ be their total income (Columns (3) and (4) present $n_k-n_{k-1}$ and $(S_k-S_{k-1})/1000$, which are the number of tax filers of each income class and their total income). Let $n$ be the ``tax unit population'', which reflects estimates for non-filers. In Table \ref{tab:PSData}, we have $n=41{,}909{,}000$. Let $p_k\coloneqq n_k/n$ be the top fractile corresponding to the $k$-th income threshold $t_k$ (Column (5) in Table \ref{tab:PSData}, in percentage point). For example, the first row ($k=37$) of Column (5) states that, in 1920, 17.32\% of tax units reported an income of at least \$1,000. The next row ($k=36$) states that 10.94\% reported at least \$2,000, and so forth. 

\begin{table}[!htb]
  \centering
  \caption{Example of the Pareto Interpolation (PI) for Tax Year 1920 using the \textit{Statistics of Income} by the Internal Revenue Service.}\label{tab:PSData}
     \begin{tabular}{rrrrrrr}
     \toprule
     (1) & (2) & (3) & (4) & (5) & (6) & (7) \\
   $k$ & Lower threshold & Tax returns & Net income & Cumulative & $b_k$ &  $a_k$ \\
    \midrule
   37 &  \$1,000  & 2,671,950 & 4,050,067 & 17.32311 & 3.27  & 1.441 \\
   36 &  \$2,000  & 2,569,316 & 6,184,543 & 10.94751 & 2.15  & 1.873 \\
   35 &  \$3,000  & 894,559 & 3,067,086 & 4.81681 & 2.23  & 1.813 \\
   34 &  \$4,000  & 442,557 & 1,972,521 & 2.68229 & 2.32  & 1.757 \\
   33 &  \$5,000  & 177,147 & 969,505 & 1.62629 & 2.48  & 1.674 \\
   32 &  \$6,000  & 112,444 & 726,362 & 1.20360 & 2.48  & 1.678 \\
   31 &  \$7,000  & 74,511 & 557,104 & 0.93529 & 2.47  & 1.682 \\
   30 &  \$8,000  & 51,211 & 434,462 & 0.75750 & 2.44  & 1.692 \\
   29 &  \$9,000  & 40,129 & 380,899 & 0.63530 & 2.41  & 1.709 \\
   28 &  \$10,000  & 29,984 & 314,400      & 0.53955 & 2.39  & 1.722 \\
   \vdots & \vdots & \vdots & \vdots & \vdots &  \vdots & \vdots\\
   2 &  \$3,000,000  & 3     & 9,218      & 0.00002 & 1.86  & 2.158 \\
   1 &  \$4,000,000  & 4     & 29,920      & 0.00001 & 1.87  & 2.149 \\
     \midrule
   & Total tax units & 41,909,000 &       &       &       & \\
    \bottomrule
    \end{tabular}%
    \caption*{\footnotesize Note: ``Net income'' is in units of \$1,000; ``Cumulative'' is in percentage points.}
\end{table}%

The Pareto interpolation (PI) method of \citet{Piketty2003} is widely used and relatively simple. Define $s_k\coloneqq S_k/n_k$ as the average income conditional of being above $t_k$. Let $b_k\coloneqq s_k/t_k$ be the ratio between the average income of taxpayers exceeding $t_k$ and the income threshold $t_k$, which is presented in Column (6) of Table \ref{tab:PSData}. If income $Y$ is Pareto distributed (in the upper tail), then for income level $y$, the cumulative distribution function (CDF) takes the form $F(y)=1-Ay^{-a}$ for some constant $A>0$ and Pareto exponent $a>1$. Therefore for any large enough\footnote{By ``large enough'', we mean that there must be enough observations. For example, if tax returns only composed the top 5\% of the tax unit population, we cannot estimate a threshold for where the top 10\% begins.} income threshold $t$, we have
\begin{equation}
    b(t)\coloneqq \frac{1}{t}\E[Y \mid Y\ge t]=\frac{1}{t}\frac{\int_t^\infty yF'(y)\diff y}{1-F(t)}=\frac{a}{a-1}. \label{eq:local_coeff}
\end{equation}
\citet{Piketty2003} refers to $b_k=s_k/t_k$ as the (local) Pareto \emph{coefficient}. When the income distribution has a Pareto upper tail, the (local) Pareto \emph{exponent} can be recovered from \eqref{eq:local_coeff} as
\begin{equation}
    a_k\coloneqq \frac{b_k}{b_k-1}=\frac{1}{1-t_k/s_k}, \label{eq:local_exp}
\end{equation}
which is presented in Column (7) of Table \ref{tab:PSData}.

With these ingredients, it becomes possible to parametrize the distribution and identify where a particular slice (\eg, top 10\%, 5\%, 1\%, 0.5\%, 0.1\%, 0.01\%) of the income distribution starts. \citet{Piketty2003} proceeds as follows to construct the top $p$ fractile income share, where $p\in (0,1]$. First, let $p_k$ be the closest proportion to $p$ directly observed in the data. For example, if $p=0.1$ (top 10\%), looking at Column (5) of Table \ref{tab:PSData}, we find that 10.94 is closest to 10, so we set $k=36$ and $p_k=0.1094$. Let $t_k$ and $a_k$ be the corresponding income threshold and local Pareto exponent. In this example, $t_k=2{,}000$ and $a_k=1.873$. Then one supposes that the income distribution is locally exactly Pareto, and therefore the CDF is
\begin{equation*}
F(y)=1-p_k(y/t_k)^{-a_k}.
\end{equation*}
The income threshold corresponding to the top $p$ fractile can be computed as
\begin{equation}
    1-p=1-p_k(y/t_k)^{-a_k}\iff y=t(p)\coloneqq t_k(p_k/p)^{1/a_k}. \label{eq:tp}
\end{equation}
Noting that the sample size is $n$, using \eqref{eq:tp}, the total income of taxpayers in the top $p$ fractile can be computed as
\begin{align}
    S(p)&\coloneqq n\int_{t(p)}^\infty yF'(y)\diff y=n\int_{t(p)}^\infty a_kp_k(y/t_k)^{-a_k}\diff y \notag \\
    &=n\frac{a_k}{a_k-1}p_kt_k^{a_k}t(p)^{1-a_k}=n\frac{a_k}{a_k-1}pt_k(p_k/p)^{1/a_k} \notag \\
    &=\underbrace{np}_\text{\# returns}\times \underbrace{b_k}_\text{Pareto coefficient}\times \underbrace{t(p)}_\text{Income threshold}. \label{eq:Yq}
\end{align}
Although the derivation is tedious, the final expression \eqref{eq:Yq} is relatively simple. To compute the total income accruing to the top $p$ fractile, all that is needed is to multiply together the number of returns in the top $p$ fractile ($np$), the local Pareto coefficient $b_k=s_k/t_k$ (which equals the ratio between the average and minimum income of those with income exceeding $t_k$), and the income threshold that defines the top $p$ fractile ($t(p)$ in \eqref{eq:tp}). Once we know $S(p)$, the top $p$ fractile income share can be computed as $S(p)/S(1)$.

Note that if $p_k=p$, \ie, the fractile in the tabular data happens to be exactly the one desired, then \eqref{eq:Yq} reduces to $S(p_k)=np_k(s_k/t_k)t_k=n_k(S_k/n_k)=S_k$, so the result is exact. This is a merit of \citet{Piketty2003}'s Pareto interpolation method as opposed to earlier ones like \citet{FeenbergPoterba1993}, where the Pareto exponent is estimated using the income thresholds and number of tax filers (but not total income), implying $S(p_k)\neq S_k$ generally.

\subsection{Maximum entropy (ME) method}\label{subsec:method_ME}

We next explain the maximum entropy method of \citet{LeeSasakiTodaWang2024JOE}. We use the same notation as in \S\ref{subsec:method_PI}. In addition, let $q_k\coloneqq (n_k-n_{k-1})/n=p_k-p_{k-1}$ be the fraction of taxpayers in income class $k$ and $y_k=(S_k-S_{k-1})/(n_k-n_{k-1})$ be their average income (which is different from the average income of \emph{all} taxpayers with income at least $t_k$, which is $s_k=S_k/n_k$).

Letting $g$ denote a generic probability density function, the definitions of $q_k$ and $y_k$ imply that
\begin{subequations}\label{eq:mcond}
\begin{align}
    \int_{t_k}^{t_{k-1}} g(y)\diff y&=q_k, \label{eq:mcond_prob}\\
    \int_{t_k}^{t_{k-1}} yg(y)\diff y&=q_ky_k \label{eq:mcond_mean}
\end{align}
\end{subequations}
for each $k \in \set{1,\dots,K}$. The idea of the maximum entropy density estimation is to find the density $f^*$ that minimizes the Kullback-Leibler (KL) divergence (with respect to the improper uniform density)
\begin{equation}
\int_{t_K}^\infty g(y)\log g(y)\diff y \label{eq:KL}
\end{equation}
subject to the moment restrictions \eqref{eq:mcond}.

To solve the maximum entropy problem, it is convenient to define the auxiliary function
\begin{equation}
    J_k(\lambda;t)\coloneqq
    \begin{cases}
        y_k\lambda-\log\left(\frac{\e^{\lambda t_{k-1}}-\e^{\lambda t_k}}{\lambda}\right), & (\lambda\neq 0)\\
        -\log \left(t_{k-1}-t_k\right). & (\lambda=0)
    \end{cases} \label{eq:Jk}
\end{equation}
\citet[Proposition 1]{LeeSasakiTodaWang2024JOE} show that $J_k$ achieves a unique maximum $\lambda_k^*$ and that the maximum entropy density is piecewise exponential given by
\begin{equation}
    f^*(y)=\begin{cases}
    \frac{q_k\lambda_k^*\e^{\lambda_k^* y}}{\e^{\lambda_k^*t_{k-1}}-\e^{\lambda_k^*t_k}}, & (\lambda_k^*\neq 0)\\
    \frac{q_k}{t_{k-1}-t_k}. & (\lambda_k^*=0)
\end{cases}\label{eq:fstar}
\end{equation}
Under some regularity conditions, \citet[Theorem 1]{LeeSasakiTodaWang2024JOE} show the strong uniform consistency of $f^*$ to the true density as the sample size $n$ tends to infinity and the income thresholds $\set{t_k}_{k=1}^K$ get finer.\footnote{In contrast, \citet{Piketty2003}'s Pareto interpolation method is ad hoc and has no mathematical proof of consistency.}

So far, we have assumed that the income thresholds $\set{t_k}_{k=1}^K$ are observed, but this is not necessary. If the researcher has access only to the cumulative numbers and incomes $\set{(n_k,S_k)}_{k=1}^K$, we could treat the income thresholds $\set{t_k}_{k=1}^K$ as unknown parameters and minimize the KL divergence further. \citet[Proposition 1]{LeeSasakiTodaWang2024JOE} show that the minimized KL divergence \eqref{eq:KL} as a function of thresholds is given by
\begin{equation}
    J^*(t_1,\dots,t_K)\coloneqq \sum_{k=1}^K q_k(J_k(\lambda_k^*;t)+\log q_k), \label{eq:Jstar}
\end{equation}
where $J_k$ is the auxiliary function in \eqref{eq:Jk}. Fixing the lower threshold $t_K$, \citet[Theorem 2]{LeeSasakiTodaWang2024JOE} show that $J^*$ achieves a unique minimum $(t_1,\dots,t_{K-1})$.\footnote{In \citet{LeeSasakiTodaWang2024JOE}, the proof of Theorem 2 is stated in Appendix B as ``Proof of Theorem 5'', which is a typographical error.} Furthermore, their proof gives the gradient and Hessian of $J^*$ explicitly, so numerically minimizing $J^*$ is straightforward.

In summary, given only the observed cumulative numbers and incomes $\set{(n_k,S_k)}_{k=1}^K$, the maximum entropy method of \citet{LeeSasakiTodaWang2024JOE} returns a nonparametric density estimate, which turns out to be a continuous piecewise exponential distribution that is highly analytically tractable. Using this density, it is straightforward to calculate top income shares at any fractile. See \url{https://github.com/alexisakira/MEtab} for the Matlab code implementing all computations.

\section{Accuracy of interpolation}\label{sec:accuracy}

\subsection{Theoretical considerations}

Both the Pareto interpolation (PI) and maximum entropy (ME) methods have their own strengths and weaknesses. Regarding ME, its strength is that the method is nonparametric. It does not rely on the Pareto assumption and thus can be applied to estimate mid-sample income shares like top 50\%. Its weakness is that because it uses only local information for estimation, the effective sample size becomes smaller and the variance may be larger. This could be an issue when estimating income shares in the far upper tail such as the top 0.01\%.

Regarding PI, its strength is its simplicity: because the method is parametric and requires the estimation of few parameters, if the parametric assumption is satisfied, the results would be likely accurate. Its weakness is that the method is susceptible to misspecification bias if the parametric assumption is violated. We can already see the misspecification from Table \ref{tab:PSData}. If income is Pareto-distributed, the local Pareto exponent $a_k$ in Column (7) should be independent of the income class $k$. However, it is decreasing in income $t_k$ in the range between top 1\% and 10\%. This means that, if $p_k>p$, then $a_k$ is larger than the true local Pareto exponent and hence inequality is underestimated. The opposite holds if $p_k<p$. This argument implies that the PI estimates are likely systematically biased in different directions depending on $p_k\gtrless p$, and the magnitude of the bias will be larger the further $p_k$ is from $p$.

\subsection{Comparison using micro-files}

The choice of Pareto interpolation (PI) comes with two potential pitfalls that must properly understood. First, in their work, \citet{PikettySaez2007} state that PI can generate non-negligible errors (p.~222, fn.~85) if there are few income classes ($k\in \set{1,\dots,K}$). By non-negligible, they apparently meant differences of ``more than 1\%'' with estimates based on micro-files (p.~222, fn.~85). The reasonable assumption being made is that micro-files are the most accurate depiction. To assess whether large differences existed is an issue, they state that they made the ``same computations for the 1966–95 period in order to compare the series estimated from Pareto interpolation with the series computed from micro-files and [\ldots] found that both series never differ by more than 1\%'' (p.~195). They do not specify the average deviation and their direction. 

\begin{figure}[!htb]
    \centering
    \includegraphics[width=0.7\linewidth]{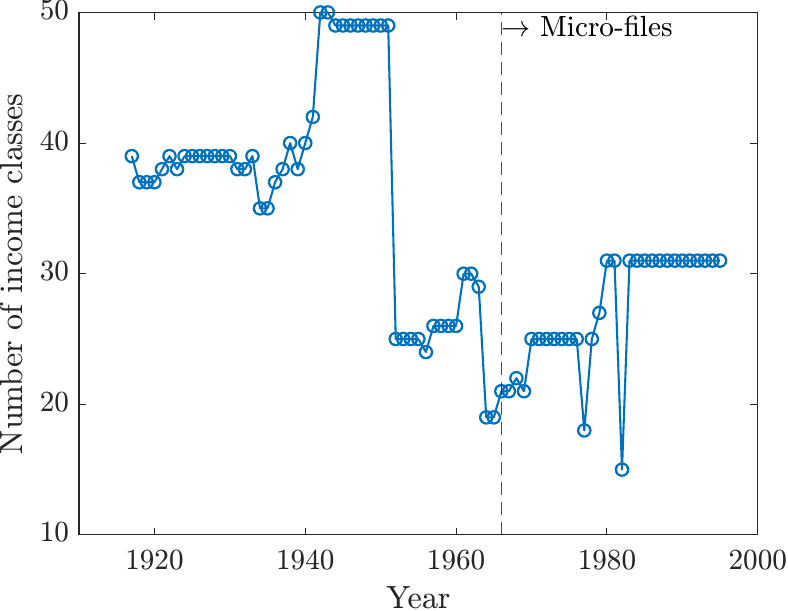}
    \caption{Number of income classes ($K$) used in \textit{Statistics of Income}, 1917 to 1995.}
    \label{fig:classes}
\end{figure}

This is not a trivial fact to point out. The reason is that the \textit{Statistics of Income} have not always reported the same number of income classes $K$ over time. The number of classes reported over time is depicted in Figure \ref{fig:classes}. In the period when \citet{PikettySaez2007} compared the micro-files results to their PI method to assess robustness (\ie, 1966 to 1995), the \textit{Statistics of Income} gradually used more and more income classes. In the early 1960s, it was using fewer than 20 income classes. By 1995, there were 31 income classes \citep[p.~24]{IRS1995}. In contrast, during the period from 1917 to 1965, the IRS was all over the place in terms of the number of classes it reported. This means that \citet{PikettySaez2007} are comparing their method with the micro-files data (that require no such interpolation) in a period of higher data quality conditions for the use of PI and are assuming that those conditions were constant. The big swings pre-1965 in the number of income classes reported are thus potentially an issue of concern. 

Second, \citet{PikettySaez2007} state that they chose the income threshold $t_k$ such that the fractile $p_k$ is as close as possible to the given $p$ they want to estimate. For example, in Table \ref{tab:PSData}, if we want to estimate $p=10\%$, we would use the income class that starts at \$2,000. However, from Column (5), we see that it is the top 10.94\% of the population. This begs the question of ``how close is close''? First, there must be a decision rule for which income class to pick. If, for example, $t_k=\$2{,}000$ had 12.5\% and $t_k=\$3{,}000$ had 8.9\% in Column (5), which would we pick? Logically, the use of the absolute distance to $p=10\%$ seems warranted. \citet{PikettySaez2007} often use this procedure, but not always. In their computation files, we found multiple years where other rules were used. For example, for the year 1917, they take the mean of the values obtained for the \$1,000 and \$2,000 classes. They do not do this for any other years. For 1940, they take the value obtained for the \$2,000 income class above which there was 14.75\% of the population. In contrast, there was 7.41\% of the tax unit population in the \$2,500 income class. The absolute distance is smaller with the latter than the former. There is thus the use of an inconsistent approach for picking the closest income fractile $p_k$. 

This might not be too problematic if, over time, the years where $p>p_k$ are followed by years where $p<p_k$ (\ie, they cancel out). The size of the differences should also cancel out over time. However, neither of these two propositions apply. There are multiple periods of many years where it is systematically the case that $p>p_k$ or $p<p_k$ and the differences tend to be larger one way. In Figure \ref{fig:pkdist} below, we show this for the top 10\% (left panel) and top 1\% (right panel) and used the distance between $p_k$ and $p$. A value less (greater) than zero implies that the income class selected has less (more) than $p$. As can be seen, from 1917 to 1943, the top 10\% tends to use income classes at and above which there is far more than 10\% of the population. From 1943 to the early 1960s, things move closer to parity. For the top 1\%, its the reverse. Most years pre-1965 tend to be for less than the top 1\%. The two last years available in that period are particularly far from 1\%.\footnote{It is worth noting that \citet{GelosoMagnessMooreSchlosser2022} were able to improve the data quality for the pre-1965 period to the differences exhibited in Figure \ref{fig:pkdist} are half as small as those of \citet{PikettySaez2007}. Their use of the PI method is thus less likely to be affected by the issue.} In other words, these are highly imprecise approximations that are being made. Moreover, the approximations have non-linear effects on the errors. Because income is distributed in a Pareto-like manner, selecting an income class where 11\% of tax units are at or above that level to represent the top 10\% will slightly underestimate inequality. In contrast, selecting an income class where 9\% of tax units are at or above that level will result in a larger error, but this time overestimating inequality. In other words, on Figure \ref{fig:pkdist}, a negative distance of the same absolute size as a positive one leads to a greater overestimation of inequality compared to the underestimation from the positive distance. Since the negative errors for both the top 10\% and 1\% tend to be concentrated in some time periods, the pattern of errors might affect trends.

\begin{figure}[!htb]
    \centering
    \includegraphics[width=0.7\linewidth]{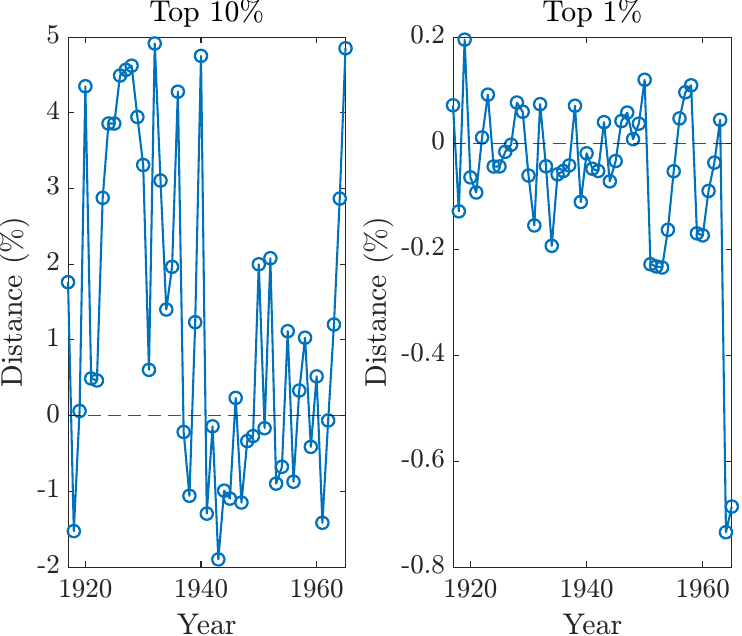}
    \caption{Distance between  $p_k$ and $p$, 1917 to 1965 (period of the use of Pareto Interpolation).}
    \label{fig:pkdist}
\end{figure}

\begin{figure}[!htb]
    \centering
    \includegraphics[width=0.7\linewidth]{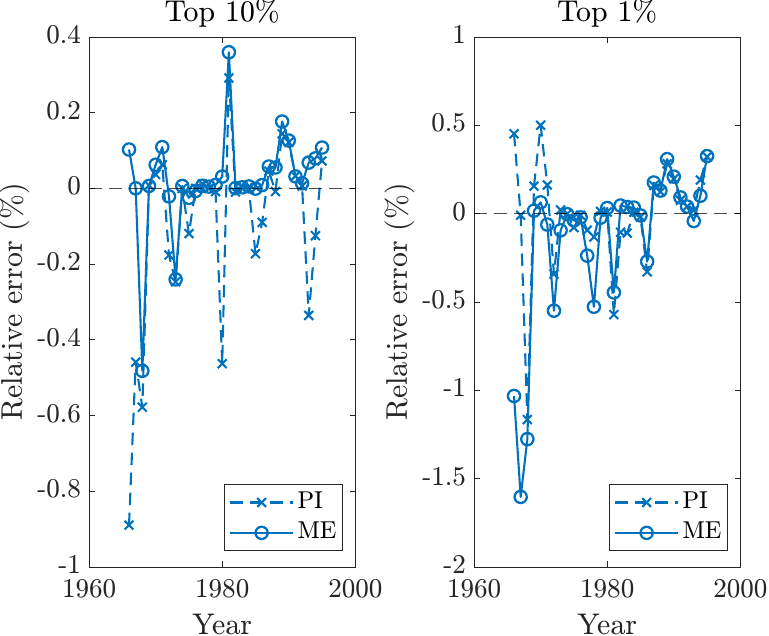}
    \caption{Relative error in top income shares, 1966 to 1995 (period micro-files are available).}
    \label{fig:relError}
\end{figure}

How well do ME and PI measure compare with the micro-files? We computed income shares using PI and ME in the 1966 to 1995 period when they can be compared with micro-files results. The idea is that the micro-files results---despite any issue one may have with them (see \citet{AutenSplinter2024})---are far closer to the ``true'' values than either PI or ME. Figure \ref{fig:relError} plots the relative error in the top income shares defined by $\hat{\theta}/\theta-1$, where $\theta$ is the true value obtained from micro-files and $\hat{\theta}$ is the estimate from either PI or ME. We see that for top 10\%, relative errors are smaller for ME than PI (mean squared error of 0.0013 and 0.0026, respectively), whereas the performance is similar for the top 1\%.

\section{Results}\label{sec:result}

\begin{figure}[!htb]
    \centering
    \includegraphics[width=0.7\linewidth]{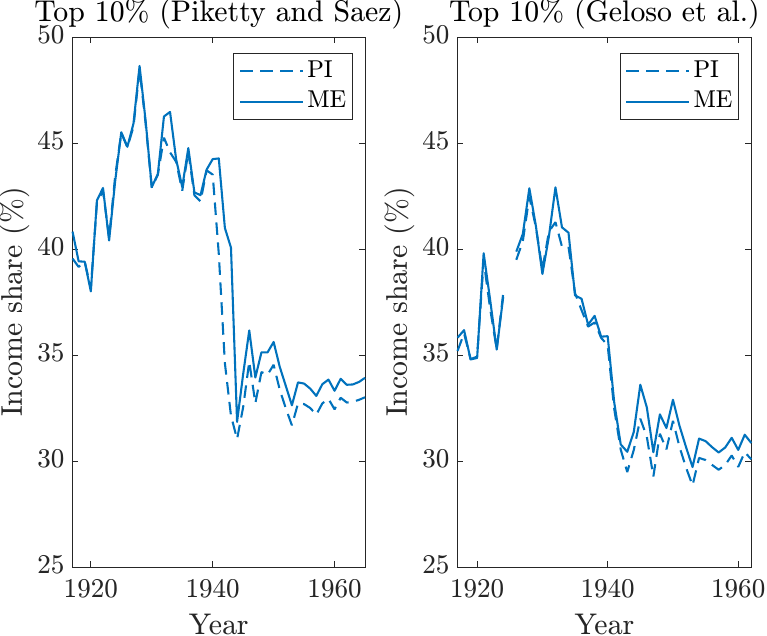}
    \caption{Income shares of the top 10\%.}
    \label{fig:top10}
\end{figure}

\begin{figure}[!htb]
    \centering
    \includegraphics[width=0.7\linewidth]{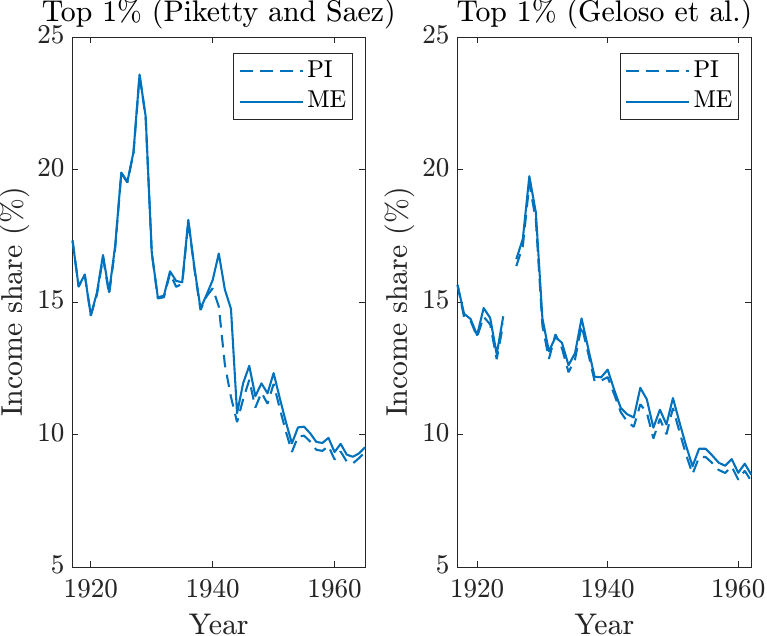}
    \caption{Income shares of the top 1\%.}
    \label{fig:top1}
\end{figure}

In Figure \ref{fig:top10}, we show the recalculated income shares for the top 10\%. Two things are obvious from above. The first is that the difference between PI and ME is relatively small in the 1920 and 1930s. This is somewhat unsurprising given the large number of income classes reported in the \textit{Statistics of Income} from 1917 to 1941 (as shown in Figure \ref{fig:classes}). This means that the difference stem largely from the distance between $p$ and $p_k$ (as shown in Figure \ref{fig:pkdist}). Second, in the early 1940s, the differences widen noticeably. This is expected because of the combined effect of falling number of income classes and the distance between $p$ and $p_k$. The differences are particularly large by the 1950s and 1960s. These statements hold regardless of whether one prefers \citet{PikettySaez2003} or \citet{GelosoMagnessMooreSchlosser2022}. They also hold when we use the top 1\% instead as in Figure \ref{fig:top1}. 

At first glance, our results appear to indicate only minor changes. However, first impressions are misleading for three reasons. First, the estimates we find are systematically higher with ME than with PI from the 1940s onwards. Identifying a higher level during that period alters the ``stylized fact'' of the U-curve of inequality in the United States. All parties involved in the debate agree on the existence of a U-curve pattern. The debate centers on how pronounced the pattern is. According to \citet{PikettySaez2003,PikettySaez2007}, the pre-1941 period was a high plateau followed by a precipitous decline in the 1940s (with a temporary peak in 1929) and a milder decline afterward. In contrast, \citet{GelosoMagnessMooreSchlosser2022} argue that the level was \textit{always} lower, with a gradual decline beginning in 1917 (with a deviation in 1929), and that a significant portion of the decline occurred during the Depression due to the wiping out of capital gains. Because our ME results for the earlier period (1917 to 1941) appear as precise as the PI results, the higher levels for the later period (1941 to 1965) imply that there was less leveling in the 1940s. This tilts the scale in favor the position of \citet{GelosoMagnessMooreSchlosser2022} who argue that the Great Depression, not the expansion of the tax system during the 1940s, was the larger contributor to the leveling. 

This particular aspect of our results aligns well with new findings that emphasize the pre-1940s as a period of greater leveling. For example, \citet{fitzsimmons2023did} recalculated the top income shares from 1921 to 1961 to account for significant regional inequalities in the cost of living that affect levels of ``real'' income inequality.\footnote{This issue is relatively minor today due to well-integrated markets within the United States \citep{slesnick2002prices,geloso2018adjusting}. However, price differences were much larger then than now \citep{coelho1979impact,haines1989state}. This means larger discrepancies between inequality estimated with ``nominal incomes'' than with ``real incomes''. } When they enacted corrections for local price level differences, they found faster leveling during the 1921 to 1941 period than previously depicted—a finding consistent with our results that the pre-1941 leveling has been underappreciated relative to the stable plateau observed from the late 1940s to the 1970s.

Second, the period going from WWII to 1975 is presented by many as the most egalitarian period in American history since the colonial period \citep{williamson1980american, goldin1992great, lindert2017unequal}. 
Our results suggest that this view is overstated; it was not as egalitarian as claimed. Since the left and right tails of the U-curve have also been shown to be lower \citep{reynolds2006income, armour2014levels, mechling2017piketty,GelosoMagnessMooreSchlosser2022,AutenSplinter2024} than depicted by \citet{PikettySaez2003,PikettySaez2007}, this pushes in favor of a different narrative of the U-curve that looks more like a ``tea saucer'' than a pronounced U-curve. 
Moreover, it is worth bearing in mind that we are using the income tax data unquestioningly. We acknowledge this is a heroic assumption. \citet{smiley2000note, geloso2020great, holcombe2021income} have pointed out that before 1943, there was no pay withholding, tax filing was voluntary, and there was little enforcement of tax obligations below the top centile, rendering the quality of the tax data quite weak. Using state-level income taxes from states with strict and aggressive enforcement, \citet{geloso2020great} revealed dramatically different distributions of income with noticeably different levels. Overall, they found that the levels of income inequality before 1943 (\ie, pre-tax withholding) were overstated by around 18\%. While there was some leveling, the lower pre-1940s levels suggest there was not as much leveling to the 1940s and 1950s as depicted by the tax data.  Our results reinforce this claim in ways that make the ``tea saucer'' analogy increasingly accurate.  

Finally, we must point out that our findings are based on pre-tax income. As of now, the only series that create inequality estimates before and after tax speak to the post-1960 period \citep{piketty2018distributional,AutenSplinter2024}. No estimates of the effects of redistribution and taxes---which would have mattered by an increasing degree after 1936 and 1943 when Social Security and pay withholding were introduced---exist before then. This is of relevance because we should not only care about the distribution of income, but the absolute living standards of those at the bottom. For example, if there is a leveling while incomes fall for everyone, this represents a normatively inferior outcome compared to a leveling where incomes rise for everyone.  The issue is that the massive expansion of the tax system in 1943 was \textit{meant} to increase the tax base by taxing a large number of (low-income) people who had never been taxed before.\footnote{Before 1940, less than 10\% of all income-earning tax units typically filed tax returns each year. This was due to a very high eligibility threshold and significant filer exemptions, which meant that most wage earners were not legally required to file. However, by the end of World War II, the percentage of tax filers surged to nearly 90\%, a level that has remained consistent ever since. This expansion occurred largely due to the lowering of the personal exemption and the increase of tax rates on everyone. For example, a single person in 1929 had a personal exemption of \$1,500 (when GDP per capita was \$858) with a tax rate of 0.375\% if in the lowest bracket (less than \$4,000). A person in the highest bracket had a tax rate of 24\% (a ratio of 64:1), In contrast, in 1943, the personal exemption was \$500 (when GDP per capita was \$1,485) with a rate of 19\% (with the lowest bracket still applying to incomes below \$4,000) \citep{IRS2002}. } If there is a great deal of leveling, holding growth constant, then an expansion of the tax base might be compensated in the after-tax income. However, because we observe \textit{less} leveling than \citet{PikettySaez2003} reported, the net improvement in living standards for those at the bottom of the income distribution will be smaller when the tax base expands to include them (and tax them at higher rates if they were already included). 
For now, this remains speculative, but it points to a direction for future research. If a method to estimate post-tax inequality before 1960 can be developed, applying the Maximum Entropy (ME) method instead of Pareto Interpolation (PI) could reveal a different pattern in the evolution of living standards for the lower segments of the population during that period.\footnote{This topic of the absolute living below the top 10\% has mostly been relegated to an afterthought since a wave of debates in the 1980s \citep{holt1977benefited,smiley1983did}. The focus is mostly on distribution. } We believe this is a vital area for further investigation.

\section{Conclusion}\label{sec:conclusion}

In this paper, we revised estimates of income inequality in America from 1917 to 1965 by improving on the method of employing tabular income tax data (from the Internal Revenue Service) to estimate distributions. The convention in the literature is to use Pareto interpolation (PI). The problem is that PI has some key vulnerabilities linked to the need to make some approximation for certain income fractiles and the need for a large number of income classes in the tabular data. We employ a new method, developed by \citet{LeeSasakiTodaWang2024JOE} and known as maximum entropy (ME), to avoid these vulnerabilities. We show that, when compared to micro-file based estimates, the maximum entropy is far more precise than the PI method---even more so in the period when the Internal Revenue Service reported few income classes in the tax data. Recalculating income shares over the period 1917 to 1965 shows that Pareto interpolation and maximum entropy yield very similar estimates until the early 1940s. However, the differences are wider from then until the 1960s. This implies that the trough of income inequality that is identified with the mid-twentieth century is not as deep. It is a higher trough. Moreover, it confirms the idea that there was relatively more leveling prior to WWII than during and after. Our results thus create a noticeably different narrative of the evolution of income inequality which should be incorporated in modern conversations regarding policy responses to the perceived rise in inequality.

Finally, we discuss some direction for future research. As discussed in the main text, because the the maximum entropy (ME) method only uses local information for estimation, it reduces the effective sample size and the variance could be large when estimating the income share of the very rich such as the top 0.01\%. To alleviate this issue, we could consider a hybrid method, say ``maximum entropy-Pareto interpolation (MEPI)'', where the ME method is used mid-sample (say the bottom 99\%) and the PI method is used for the top 1\%. For this purpose, we could use the efficient minimum distance estimation of the Pareto exponent proposed by \citet{TodaWang2021JAE}. Similarly, we also point out that our method should constitute an encouragement to extend estimates of post-tax income distribution before the 1960s. 

\printbibliography

\newpage
\appendix

\section{Table of new estimates}

\begin{longtable}{ccccccc}
\caption{Revised income shares using \citet{PikettySaez2003}.} \\
\hline
Year & P90-100 & P95-100 & P99-100 & P99.5-100 & P99.9-100 & P99.99-100 \\
\hline
\endfirsthead
\caption{(continued)} \\
\hline
Year & P90-100 & P95-100 & P99-100 & P99.5-100 & P99.9-100 & P99.99-100 \\
\hline
\endhead
\hline \\
\endfoot
1917  & 40.82 & 29.85 & 17.32 & 14.00 & 8.20  & 3.29 \\
1918  & 39.42 & 28.80 & 15.58 & 12.14 & 6.56  & 2.40 \\
1919  & 39.40 & 29.47 & 16.04 & 12.35 & 6.42  & 2.24 \\
1920  & 38.01 & 27.69 & 14.50 & 10.89 & 5.24  & 1.63 \\
1921  & 42.29 & 30.21 & 15.35 & 11.49 & 5.50  & 1.66 \\
1922  & 42.88 & 31.38 & 16.77 & 12.84 & 6.52  & 2.23 \\
1923  & 40.40 & 29.26 & 15.38 & 11.70 & 5.80  & 1.96 \\
1924  & 43.22 & 31.55 & 17.13 & 13.17 & 6.67  & 2.28 \\
1925  & 45.51 & 34.35 & 19.89 & 15.58 & 8.37  & 3.25 \\
1926  & 44.82 & 33.94 & 19.54 & 15.26 & 8.30  & 3.30 \\
1927  & 45.92 & 35.05 & 20.66 & 16.31 & 9.09  & 3.69 \\
1928  & 48.64 & 37.97 & 23.59 & 19.11 & 11.37 & 4.95 \\
1929  & 46.03 & 35.89 & 22.00 & 17.79 & 10.74 & 4.91 \\
1930  & 42.92 & 31.42 & 16.88 & 12.94 & 6.93  & 2.79 \\
1931  & 43.54 & 30.72 & 15.17 & 11.32 & 5.77  & 2.20 \\
1932  & 46.25 & 31.78 & 15.25 & 11.39 & 5.85  & 1.95 \\
1933  & 46.47 & 32.77 & 16.16 & 12.24 & 6.49  & 2.30 \\
1934  & 44.23 & 32.41 & 15.80 & 11.66 & 5.77  & 1.92 \\
1935  & 42.93 & 30.85 & 15.75 & 11.85 & 5.94  & 2.00 \\
1936  & 44.76 & 33.08 & 18.10 & 13.84 & 6.98  & 2.31 \\
1937  & 42.67 & 30.95 & 16.30 & 12.32 & 6.11  & 2.01 \\
1938  & 42.54 & 29.89 & 14.76 & 10.92 & 5.32  & 1.85 \\
1939  & 43.75 & 30.90 & 15.28 & 11.32 & 5.45  & 1.77 \\
1940  & 44.24 & 31.31 & 15.83 & 11.76 & 5.65  & 1.85 \\
1941  & 44.27 & 32.07 & 16.83 & 12.50 & 5.96  & 1.92 \\
1942  & 40.99 & 29.67 & 15.47 & 11.41 & 5.26  & 1.60 \\
1943  & 40.07 & 29.17 & 14.74 & 10.70 & 4.75  & 1.39 \\
1944  & 31.85 & 22.17 & 10.80 & 7.82  & 3.46  & 1.00 \\
1945  & 34.10 & 24.33 & 11.94 & 8.55  & 3.72  & 1.07 \\
1946  & 36.16 & 26.10 & 12.59 & 8.95  & 3.87  & 1.15 \\
1947  & 33.94 & 24.19 & 11.45 & 8.12  & 3.52  & 1.06 \\
1948  & 35.13 & 24.92 & 11.93 & 8.55  & 3.75  & 1.11 \\
1949  & 35.13 & 24.61 & 11.56 & 8.24  & 3.61  & 1.09 \\
1950  & 35.62 & 25.33 & 12.32 & 8.88  & 3.98  & 1.25 \\
1951  & 34.45 & 24.10 & 11.38 & 8.10  & 3.53  & 1.05 \\
1952  & 33.54 & 23.05 & 10.48 & 7.37  & 3.10  & 0.90 \\
1953  & 32.63 & 22.05 & 9.67  & 6.73  & 2.79  & 0.81 \\
1954  & 33.71 & 23.00 & 10.27 & 7.19  & 3.04  & 0.92 \\
1955  & 33.65 & 23.02 & 10.30 & 7.21  & 3.10  & 0.99 \\
1956  & 33.42 & 22.75 & 10.05 & 6.99  & 2.94  & 0.92 \\
1957  & 33.07 & 22.37 & 9.72  & 6.71  & 2.78  & 0.84 \\
1958  & 33.63 & 22.61 & 9.68  & 6.67  & 2.76  & 0.84 \\
1959  & 33.84 & 22.82 & 9.87  & 6.83  & 2.84  & 0.88 \\
1960  & 33.32 & 22.23 & 9.34  & 6.40  & 2.68  & 0.86 \\
1961  & 33.88 & 22.71 & 9.65  & 6.66  & 2.87  & 0.96 \\
1962  & 33.60 & 22.34 & 9.24  & 6.29  & 2.60  & 0.84 \\
1963  & 33.61 & 22.31 & 9.16  & 6.22  & 2.57  & 0.83 \\
1964  & 33.73 & 22.51 & 9.29  & 6.35  & 2.71  & 0.83 \\
1965  & 33.92 & 22.75 & 9.52  & 6.59  & 2.88  & 0.92 \\
\end{longtable}%
\newpage
\begin{longtable}{ccccccc}
\caption{Revised Income Shares using \citet{GelosoMagnessMooreSchlosser2022}.} \\
\hline
Year & P90-100 & P95-100 & P99-100 & P99.5-100 & P99.9-100 & P99.99-100 \\
\hline
\endfirsthead
\caption{(continued)} \\
\hline
Year & P90-100 & P95-100 & P99-100 & P99.5-100 & P99.9-100 & P99.99-100 \\
\hline
\endhead
\hline \\
\endfoot

    1917  & 35.82 & 26.66 & 15.65 & 12.65 & 7.41  & 2.98 \\
    1918  & 36.18 & 26.76 & 14.56 & 11.31 & 6.14  & 2.29 \\
    1919  & 34.79 & 26.35 & 14.36 & 11.02 & 5.80  & 2.08 \\
    1920  & 34.86 & 25.80 & 13.75 & 10.48 & 5.15  & 1.65 \\
    1921  & 39.80 & 28.84 & 14.78 & 11.19 & 5.45  & 1.71 \\
    1922  & 37.59 & 27.49 & 14.40 & 11.04 & 5.64  & 1.95 \\
    1923  & 35.26 & 25.40 & 13.03 & 9.94  & 4.94  & 1.68 \\
    1924  & 37.75 & 27.31 & 14.46 & 11.12 & 5.62  & 1.94 \\
    1925  &   -    & -      &    -   &   -    &   -    & - \\
    1926  & 39.88 & 29.87 & 16.62 & 12.92 & 6.98  & 2.76 \\
    1927  & 40.71 & 30.64 & 17.38 & 13.68 & 7.59  & 3.06 \\
    1928  & 42.87 & 32.96 & 19.75 & 15.94 & 9.41  & 4.06 \\
    1929  & 41.10 & 31.39 & 18.37 & 14.79 & 8.85  & 3.99 \\
    1930  & 38.82 & 27.86 & 14.38 & 11.00 & 5.86  & 2.35 \\
    1931  & 40.60 & 27.99 & 13.14 & 9.80  & 4.98  & 1.90 \\
    1932  & 42.91 & 29.33 & 13.68 & 10.01 & 5.12  & 1.73 \\
    1933  & 41.02 & 28.19 & 13.46 & 10.11 & 5.33  & 1.90 \\
    1934  & 40.77 & 29.94 & 12.62 & 9.42  & 4.65  & 1.56 \\
    1935  & 37.81 & 26.53 & 13.07 & 9.82  & 4.92  & 1.66 \\
    1936  & 37.66 & 27.23 & 14.38 & 10.99 & 5.55  & 1.85 \\
    1937  & 36.43 & 25.96 & 13.27 & 10.04 & 5.01  & 1.69 \\
    1938  & 36.85 & 25.43 & 12.18 & 9.01  & 4.40  & 1.54 \\
    1939  & 35.87 & 25.15 & 12.16 & 9.03  & 4.37  & 1.44 \\
    1940  & 35.89 & 25.26 & 12.45 & 9.26  & 4.47  & 1.49 \\
    1941  & 32.77 & 23.17 & 11.69 & 8.71  & 4.20  & 1.37 \\
    1942  & 30.78 & 21.76 & 11.00 & 8.18  & 3.84  & 1.18 \\
    1943  & 30.44 & 21.46 & 10.76 & 7.92  & 3.61  & 1.07 \\
    1944  & 31.38 & 21.84 & 10.64 & 7.70  & 3.41  & 0.99 \\
    1945  & 33.60 & 23.97 & 11.76 & 8.42  & 3.66  & 1.06 \\
    1946  & 32.53 & 23.48 & 11.33 & 8.05  & 3.48  & 1.03 \\
    1947  & 30.41 & 21.68 & 10.26 & 7.28  & 3.16  & 0.95 \\
    1948  & 32.20 & 22.85 & 10.94 & 7.84  & 3.44  & 1.01 \\
    1949  & 31.56 & 22.11 & 10.38 & 7.41  & 3.24  & 0.98 \\
    1950  & 32.89 & 23.39 & 11.38 & 8.20  & 3.68  & 1.16 \\
    1951  & 31.67 & 22.16 & 10.46 & 7.45  & 3.24  & 0.97 \\
    1952  & 30.66 & 21.07 & 9.58  & 6.74  & 2.84  & 0.82 \\
    1953  & 29.72 & 20.08 & 8.81  & 6.12  & 2.54  & 0.74 \\
    1954  & 31.06 & 21.19 & 9.46  & 6.62  & 2.80  & 0.84 \\
    1955  & 30.94 & 21.16 & 9.47  & 6.63  & 2.85  & 0.91 \\
    1956  & 30.65 & 20.87 & 9.22  & 6.41  & 2.69  & 0.84 \\
    1957  & 30.40 & 20.56 & 8.94  & 6.17  & 2.56  & 0.77 \\
    1958  & 30.64 & 20.60 & 8.82  & 6.08  & 2.52  & 0.77 \\
    1959  & 31.10 & 20.97 & 9.07  & 6.28  & 2.61  & 0.81 \\
    1960  & 30.53 & 20.37 & 8.56  & 5.87  & 2.46  & 0.79 \\
    1961  & 31.24 & 20.94 & 8.90  & 6.14  & 2.65  & 0.88 \\
    1962  & 30.85 & 20.51 & 8.49  & 5.77  & 2.39  & 0.77 \\
\end{longtable}%

\end{document}